\documentclass{aa}  
\usepackage{color}
\usepackage{ulem}
\usepackage{soul}

\usepackage{graphicx}
\usepackage{txfonts}
\begin{document}

   \title{Magnetic properties on the boundary of an evolving pore }


   \author{  M. Garc\'ia-Rivas  \inst{1,2},
             J. Jur\v c\'ak      \inst{1}
            \and
            N. Bello Gonz\'alez \inst{3}        
          }

   \institute{  Astronomical Institute of the Czech Academy of Sciences, Fri\v cova 298, 25165 Ond\v rejov, Czech Republic \\
                \email{rivas@asu.cas.cz}    
                \and
                Astronomical Institute, Charles University, V Hole\v sovick\'ach 2, 18000 Praha, Czech Republic           
                \and
                Leibniz-Institut f\"ur Sonnenphysik (KIS), Sch\"oneckstrasse 6, 79104 Freiburg, Germany
             }

   \date{Received 12 October 2020 / Accepted 8 February 2021}

 
  \abstract
   {Analyses of the magnetic properties on umbrae boundaries have led to the Jur\v c\'ak criterion, which states that umbra-penumbra boundaries in stable sunspots are equally defined by a constant value of the vertical magnetic field, $B_\mathrm{ver}^\mathrm{crit}$, and by a 50\% continuum intensity of the quiet Sun, $I_\mathrm{QS}$. Umbrae with vertical magnetic fields stronger than $B_\mathrm{ver}^\mathrm{crit}$ are stable, whereas umbrae with vertical magnetic fields weaker than $B_\mathrm{ver}^\mathrm{crit}$ are unstable and prone to vanishing.}
   {We aim to investigate the existence of a critical value of the vertical magnetic field on a pore boundary and its role in the evolution of the magnetic structure.}
   {We analysed SDO/HMI vector field maps corrected for scattered light and with a temporal cadence of 12 min during a 26.5-hour period. A continuum intensity threshold ($I_\mathrm{c}~=~0.55\,I_\mathrm{QS}$) is used to define the pore boundary and we study the temporal evolution of the magnetic properties there.}
   { We observe well-defined stages in the pore evolution:  (1) during the initial formation phase, total magnetic field strength ($B$) and vertical magnetic field ($B_\mathrm{ver}$) increase to their maximum values of $\sim1920$~G and $\sim1730$~G, respectively;  (2) then the pore reaches a stable phase; (3) in a second formation phase, the pore undergoes a rapid growth in terms of size, along with a decrease in $B$ and $B_\mathrm{ver}$ on its boundary. In the newly formed area of the pore, $B_\mathrm{ver}$ remains mostly below 1731~G and $B$ remains everywhere below 1921~G; (4) ultimately, pore decay starts. We find overall that pore areas with $B_\mathrm{ver} < 1731$~G, or equivalently $B < 1921$~G, disintegrate faster than regions that fulfil this criteria.}
   {We find that the most stable regions of the pore, similarly to the case of umbral boundaries, are defined by a critical value of the vertical component of the magnetic field that is comparable to that found in stable sunspots. In addition, in this case study, the same pore areas can be similarly well-defined by a critical value of the total magnetic field strength.}


   \maketitle
%

\section{Introduction}

When observed by telescope, the solar surface appears covered in a granular pattern. These granules are the tops of convective cells transporting hot plasma from the solar interior to the solar surface, thus heating the photospheric layer of the solar atmosphere. \citet{Hale:1908} discovered that sunspots are caused by strong magnetic fields. These strong magnetic fields inhibit the convective motions in the sub-surface layers of the sun resulting in cooler (darker) spots on the solar surface. There are also smaller dark structures observed on the solar disc called pores. Since these discoveries were made, an  ongoing effort has been underway to investigate the properties of magnetic fields and dynamics in sunspots and pores \citep[see e.g.][]{Keppens:1996, Solanki:2003, Schlichenmaier:2009, Borrero:2011}.

Sunspots are structures composed of two distinctive areas: umbra and penumbra. Magnetic fields within umbrae are stronger and more vertical compared to the penumbra and the magneto-convection within them leads to the formation of umbral dots \citep[e.g.][]{Schussler_etal2006,Ortiz_etal2010}. Magnetic fields within penumbral regions are weaker and more horizontal than in umbrae, and magneto-convection in penumbrae leads to highly elongated cells that form the penumbral filaments \citep[e.g.][]{Rempel2011, Tiwari_etal2013}. Pores, on the other hand, are composed of only one distinctive area that morphologically resembles sunspot umbrae. The non-existence of penumbra around pores has been explained by a simple magnetic flux tube model with a prevailing vertical field \citep[e.g.][]{Simon_etal1970}. Subsequent high spatial resolution observations showed that some pores contain fine bright features, such as light bridges or umbral dots, that reveal magneto-convection \citep[e.g.][]{Sobotka_etal1999, Hirzberger2003, Giordano_etal2008, Sobotka_etal2009}. 

Since there were no known magnetic properties to define them, umbra-penumbra (UP) boundaries were traditionally defined by a continuum intensity ($I_\mathrm{c}$) threshold. In \cite{Jurcak2011}, the spectropolarimetric analysis of ten sunspots provided the first hint at the invariance of $B_\mathrm{ver}$ on UP boundaries. In an extended analysis of 79 active regions observed with Hinode/SOT-SP, \cite{Jurcak_etal2018} gave statistical proof of the magnetic nature of stable UP boundaries with a critical vertical magnetic field of $B_\mathrm{ver}^\mathrm{crit}~=~1867~\pm~18$~G. The analysis of a long-lived sunspot during its stable phase observed with SDO/HMI resulted in a value of $B_\mathrm{ver}^\mathrm{crit}~=~1693~\pm~15$~G \citep{Schmassmann_etal2018}. The difference between \citet{Jurcak_etal2018} and \citet{Schmassmann_etal2018} has been explained by the different instrument resolutions and analysis methods employed.

\begin{figure*}
\sidecaption
\includegraphics[width=0.7\linewidth]{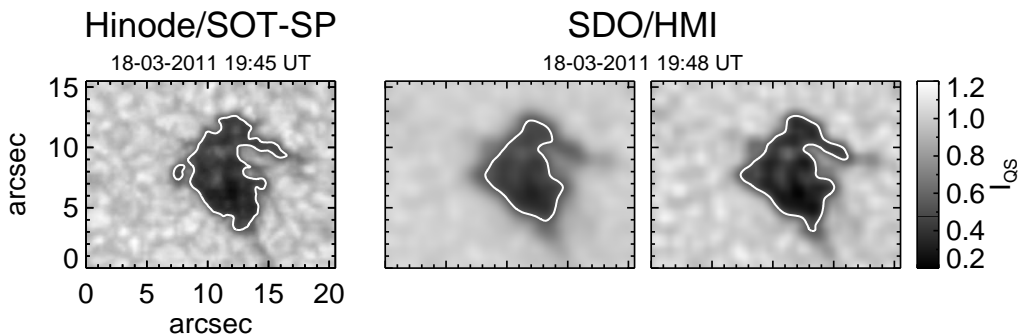}
\caption{Comparison of continuum intensity maps of the pore on 18 March 2011 observed with Hinode/SOT-SP at 19:45~UT (left) and regular (middle) and deconvolved (right) HMI/SDO maps at 19:48~UT. Isocontours at $I_\mathrm{c}~=~0.55\,I_\mathrm{QS}$ are marked with a white curve.}
\label{fig:hinode_to_hmi}
\end{figure*}

Additionally, unstable sunspots have also been investigated. During the formation of a sunspot, it was found that penumbra partially colonises umbral areas; over several hours, the UP boundary migrates to regions with a stronger $B_\mathrm{ver}$ until $B_\mathrm{ver}^\mathrm{crit}$ is reached and the position of the boundary stabilises \citep{Jurcak_etal2015}. On the other hand, during the decay of a sunspot, $B_\mathrm{ver}$ is neither constant nor strong enough on the UP boundary; the umbra is prone to be disintegrated by convection or magnetic diffusion and this process takes place over several days \citep{Benko_etal2018}. 

The previously described observational studies have led to the so-called Jur\v c\'ak criterion \citep[first introduced in][]{Schmassmann_etal2018}. This empirical law not only states that stable UP boundaries can be described either by 50\% continuum intensity of the quiet Sun, $I_\mathrm{c}=0.5\,I_\mathrm{QS}$, or by a critical vertical magnetic field $B_\mathrm{ver}^\mathrm{crit}$, it also states that $B_\mathrm{ver}\geq B_\mathrm{ver}^\mathrm{crit}$ only allow for umbral modes of magneto-convection, while umbral areas with $B_\mathrm{ver}~<~B_\mathrm{ver}^\mathrm{crit}$ are unstable and prone to vanish under more vigorous convective modes. 

These observational results are in agreement with simplified theoretical studies of convective motions in the presence of magnetic fields. In \citeyear{Chandrasekhar1961}, Chandrasekhar derived the effects of inclined magnetic fields on convection; he showed that the convective mode is sensitive to the vertical magnetic field, $B_\mathrm{ver}$, whereas the convective cells shape is sensitive to the horizontal component, $B_{hor}$. This was further supported by the theoretical study of \citet{Gough:1966} on the stabilising effect against overturning convection of the vertical component of the magnetic field, thus expanding the Schwarzschild criterion for convective instability. Stability within this criterion is defined as the hindering of an instability or perturbation to develop overturning convection: Schwarzschild stated that convection sets in when the temperature gradient is steeper than the adiabatic gradient ($\nabla~>~\nabla_\mathrm{ad}$, where $\nabla \equiv d ~\mathrm{ln}T / d~ \mathrm{ln} p$, and \textit{T}, \textit{p} are the local temperature and pressure, respectively). \citet{Gough:1966} derived an extra parameter to expand the Schwarzchild criterion; this new constraint to keep stability depends not only on the pressure, but also on the vertical component of the magnetic field,  $B_\mathrm{ver}$. This is a simplistic criterion that does not take into account other processes that influence the triggering of convective motions, such as rotation.

Analyses of the relevant observations focusing on the role of $B_\mathrm{ver}$ \citep{Jurcak2011, Jurcak_etal2015, Jurcak_etal2017, Jurcak_etal2018, Schmassmann_etal2018, Benko_etal2018, Lindner:2020} suggest the validity of the theoretical studies by \citet{Chandrasekhar1961} and \citet{Gough:1966}. Although it is not specifically mentioned in any of these observational studies, we understand the $B_\mathrm{ver}^\mathrm{crit}$ value as the photospheric counterpart of the critical vertical field stabilising the sub-photospheric layers against more vigorous modes of (magneto-) convection.

Magnetic properties have so far been studied exclusively on umbral boundaries. In this paper, we investigate the magnetic properties on the boundary of an evolving pore to study the applicability of the Jur\v c\'ak criterion to the stability of pores against more vigorous modes of magneto-convection.

\section{Data analysis} \label{sec:analysis}

In this study, we analyse a pore observed in the active region NOAA 11175 from 09:24~UT on 18 March 2011 ($\mu$~=~0.94, with $\mu=cos \theta$, where $\theta$ is the heliocentric angle) to 12:00~UT on 19 March 2011 ($\mu$~=~0.90). 

An analysis of the magnetic properties of an evolving pore requires spectropolarimetric data with a temporal cadence that is much higher than the evolutionary timescales of a pore, that is, we require abundant measures during the lifetime of a pore, which typically spans from hours to days. We also require observations with enough spatial resolution to resolve the magnetic structure. 

The Helioseismic and Magnetic Imager \citep[HMI][]{HMI_Schou2012} on-board the Solar Dynamics Observatory \citep[SDO][]{SDO_Presnell2012} fulfils our requirements. HMI provides full-disc Stokes parameters every 12 minutes, with a pixel scale of $\approx$$0,5''$ and stable conditions. However, the observations are limited by a moderate spatial resolution of $\approx$$1''$. As in any optical setting, spatial resolution is diminished by the convolution of a real image with the point spread function (PSF) of the instrument. Therefore, a proper characterisation of the HMI's PSF could improve the quality of the images. \cite{HMI_PSF_dcon} have modelled HMI's PSF as an Airy function convolved with a Lorentzian; they correct images from both large-scale wave-front errors and long-distance scattering based on tests prior to the launch of SDO \citep{HMI_study_beforelaunch}, and in-flight Venus-transit and Lunar-eclipse observations. Deconvolved HMI maps, that is, regular HMI maps corrected for scattered light using the described PSF model, are in qualitative agreement with sub-arcsecond spatial resolution observations \citep[e.g.][]{HMI_dcon_Norton}. Deconvolved maps exhibit a better spatial resolution, which implies a higher continuum intensity contrast in granulation and darker magnetic structures such as umbrae.

Figure~\ref{fig:hinode_to_hmi} compares a sub-arcsecond spatial resolution scan from the spectropolarimeter on-board Hinode satellite \citep{SOT_Tsuneta2008, Hinode_Kosugi2007} to regular and deconvolved HMI maps. The deconvolved image not only exhibits a better spatial resolution, but a more consistent intensity map. Consequently, our analysis of the temporal evolution of a pore was performed using deconvolved HMI maps, specifically  HMI data sets \texttt{hmi.B\_720s\_dconS} and \texttt{hmi.Ic\_720s\_dconS} kindly processed for us by A. Norton. Henceforth, every time we mention HMI maps, we are referring to the HMI deconvolved maps.

The HMI measures the Stokes parameters at six wavelength positions along the \ion{Fe}{I}~617.3\ nm line on the full-solar disc. The HMI Vector Magnetic Field Pipeline \citep{HMI_Pipeline_Hoeksema2014} automatically computes the photospheric vector magnetic field using the Very Fast Inversion of the Stokes Vector code \citep[VFISV,][]{VFISV_Borrero2011,VFISV_Centeno2014}. For our purposes, we use three inferred parameters:\ magnetic field strength, inclination, and azimuth, which are in a line-of-sight reference frame (LOS). We transfer the magnetic field vector to the local reference frame (LRF) by solving the azimuth ambiguity first. In pores, we can use the solution provided by the ME0 method, a variation of the Minimum-Energy method \citep{disambiguation_metcalf,disambiguation_leka}, available for all HMI data sets. In $180^{\circ}$-rotated HMI maps, $180^{\circ}$ must be added to the azimuth. The transformation of the angular parameters of the magnetic field from LOS to LRF is then performed with the routine \texttt{r\_frame\_sphduo.pro} from the AZAM package \citep{Lites_etal1995}. Thus, we obtain the magnetic parameters necessary for our analysis: magnetic field strength ($B$), magnetic field inclination in LRF ($\gamma$), and the vertical component of the magnetic field ($B_\mathrm{ver}=B\,cos\gamma$). The 24-hour orbital induced variation on the magnetic parameters was calculated on the umbral boundary of a stable sunspot during a period of about ten days by \cite{Schmassmann_etal2018}. They found that $B$ and $B_\mathrm{ver}$ oscillated less than $20\, G$ and $\gamma$ oscillated $0.2\,^{\circ}$. Our analysis of the pore does not allow us to study the orbital induced variations in full, however, based on the analysis by \cite{Schmassmann_etal2018}, the difference is almost negligible compared to the evolutionary changes in $B$, $B_\mathrm{ver}$, and $\gamma$.

Continuum intensity maps are normalised to local quiet Sun intensity ($I_\mathrm{QS}$). In order to discard the centre-to-limb and orbital induced variations, local quiet Sun intensity is calculated as the mean continuum intensity of a quiet Sun sub-region close to the studied pore for each frame. 

For the purpose of defining the boundary of the pore in terms of the continuum intensity, we investigate isocontours in the range $(0.4-0.6)\,I_\mathrm{QS}$. In order to  carry out a systematic study of the evolution of magnetic properties on the pore boundary, we need to select a unique intensity isocontour. We are aware that in some cases a unique isocontour does not completely match the boundary of the pore. The dimmer regions of the pore that are not included in the selected isocontour evolve more rapidly than the main structure of the pore. Therefore, we believe that using a higher value of intensity threshold would only contaminate our results by adding rapid variations of the magnetic properties. The chosen isocontour at $I_\mathrm{c}=0.55\,I_\mathrm{QS}$ provides a good visual match to the pore's boundary while avoiding the introduction of strong variations on the magnetic properties.

Due to the difference between the formation heights of the continuum intensity and the derived magnetic properties \citep[the \ion{Fe}{I}~617.3\ nm line is more sensitive to magnetic fields around $\log \tau = -1$,][]{FeI_6173_Nazaret2009}, there are projection effects that can influence our analysis. In the case of sunspots, observations in areas away from the disc centre resulted in shifts between intensity and magnetic isocontours \citep{Jurcak_etal2018, Schmassmann_etal2018} or analogically systematic variation of magnetic properties along intensity boundaries \citep{Jurcak2011}. \cite{Schmassmann_etal2018} calculated a shift of 1.3 pixel on a UP boundary at the limb, therefore, the projection effects are negligible in our case study, since the observed pore is located close to the disc centre ($\mu \in [0.90 - 0.94]$).

\begin{figure}
 \centering
  \includegraphics[width=0.9\linewidth] {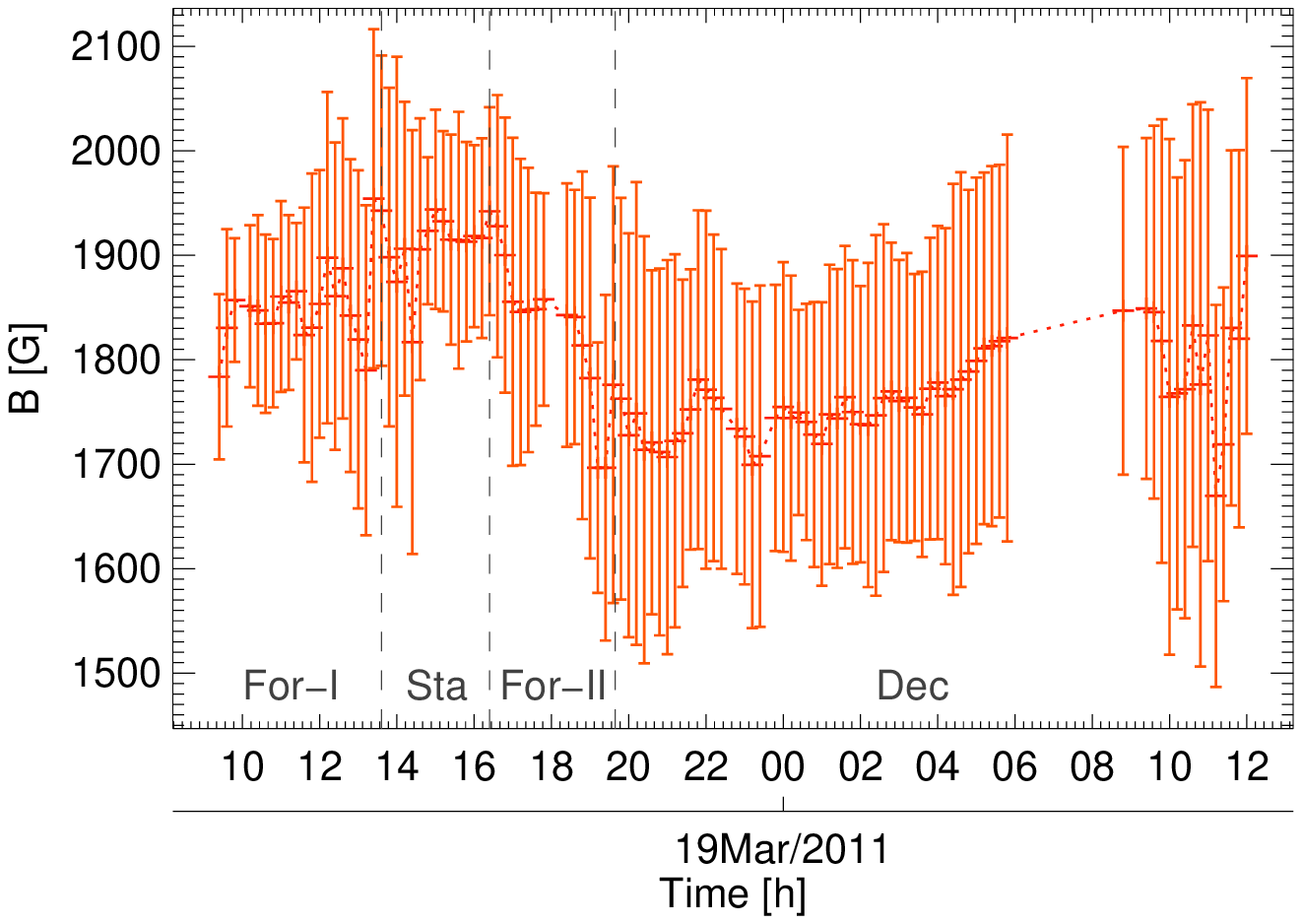}
  \includegraphics[width=0.9\hsize] {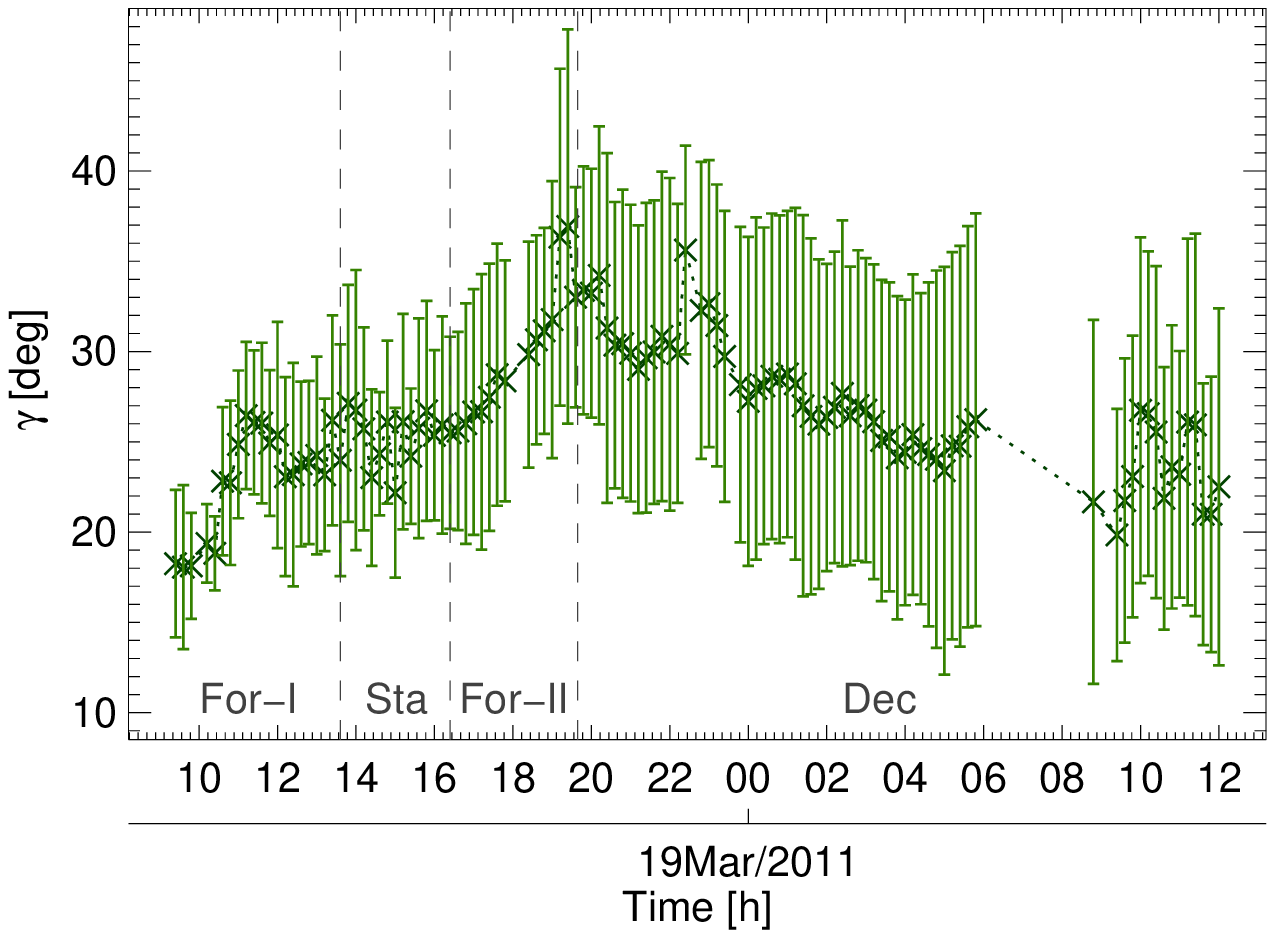}
  \includegraphics[width=0.9\hsize] {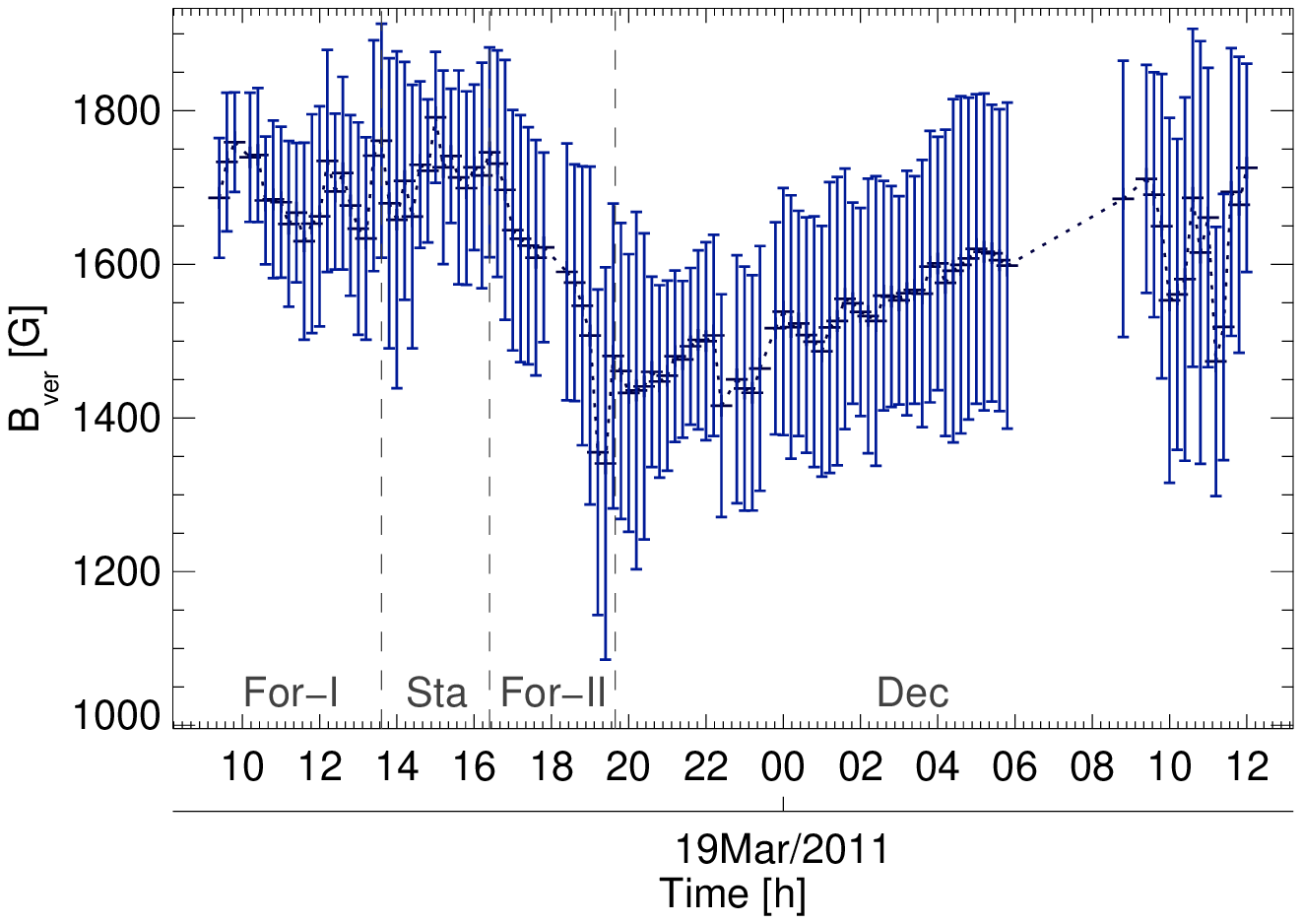}
  \caption{Temporal evolution of the averaged magnetic parameters on the boundary of the pore ($I_c=0.55\,I_\mathrm{QS}$). From top to bottom, we show $B$, $\gamma$, and $B_\mathrm{ver}$. The uncertainties are given by the standard deviation of these physical parameters for each observation. The plot is divided in evolutionary stages: first period of formation (For-I), stability (Sta), second period of formation (For-II), and decay (Dec). }

  \label{fig:bver_evolution}
\end{figure}

\begin{figure}
\centering
{\includegraphics[width=1\linewidth]{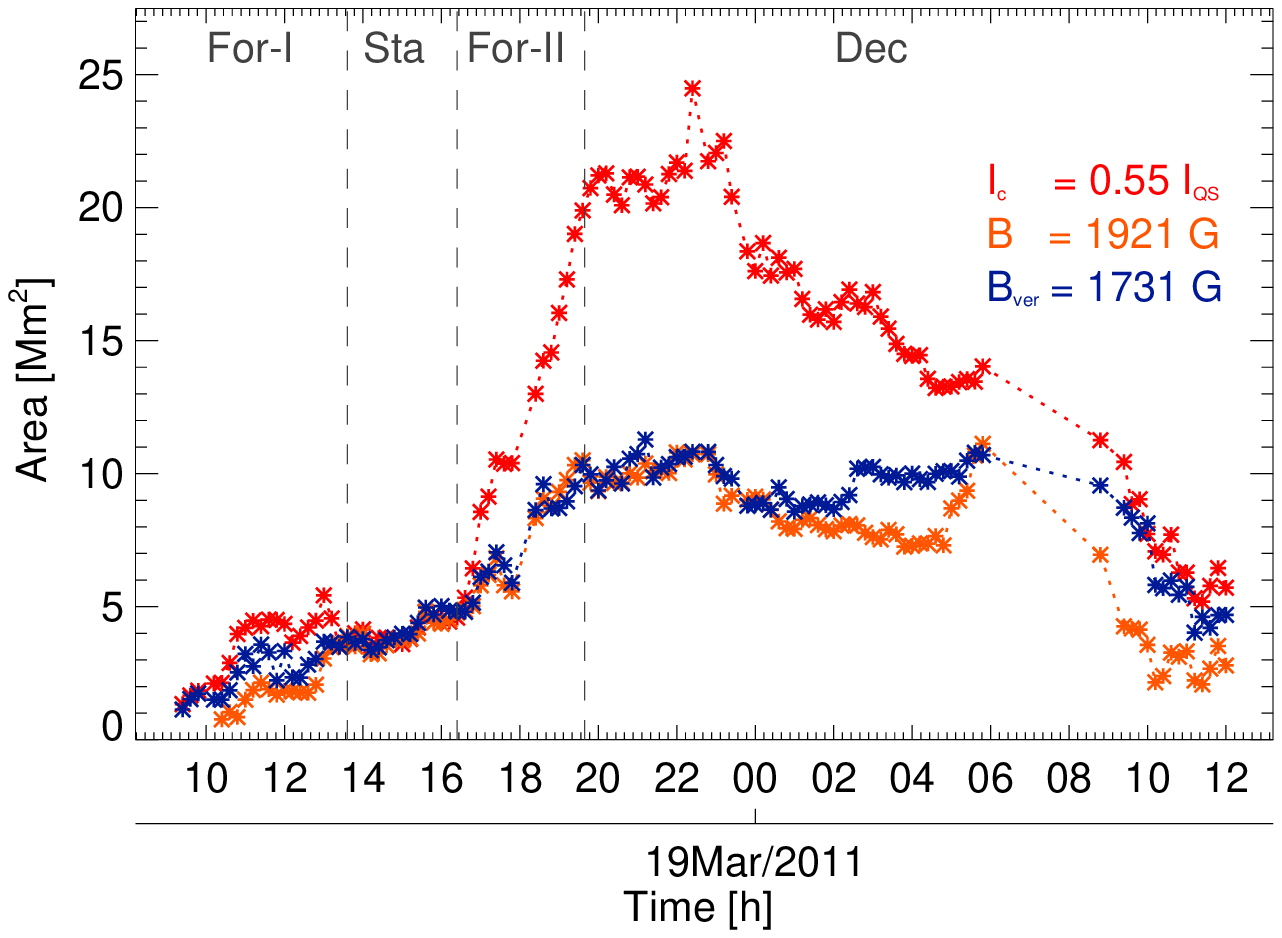}}
{\includegraphics[width=1\linewidth]{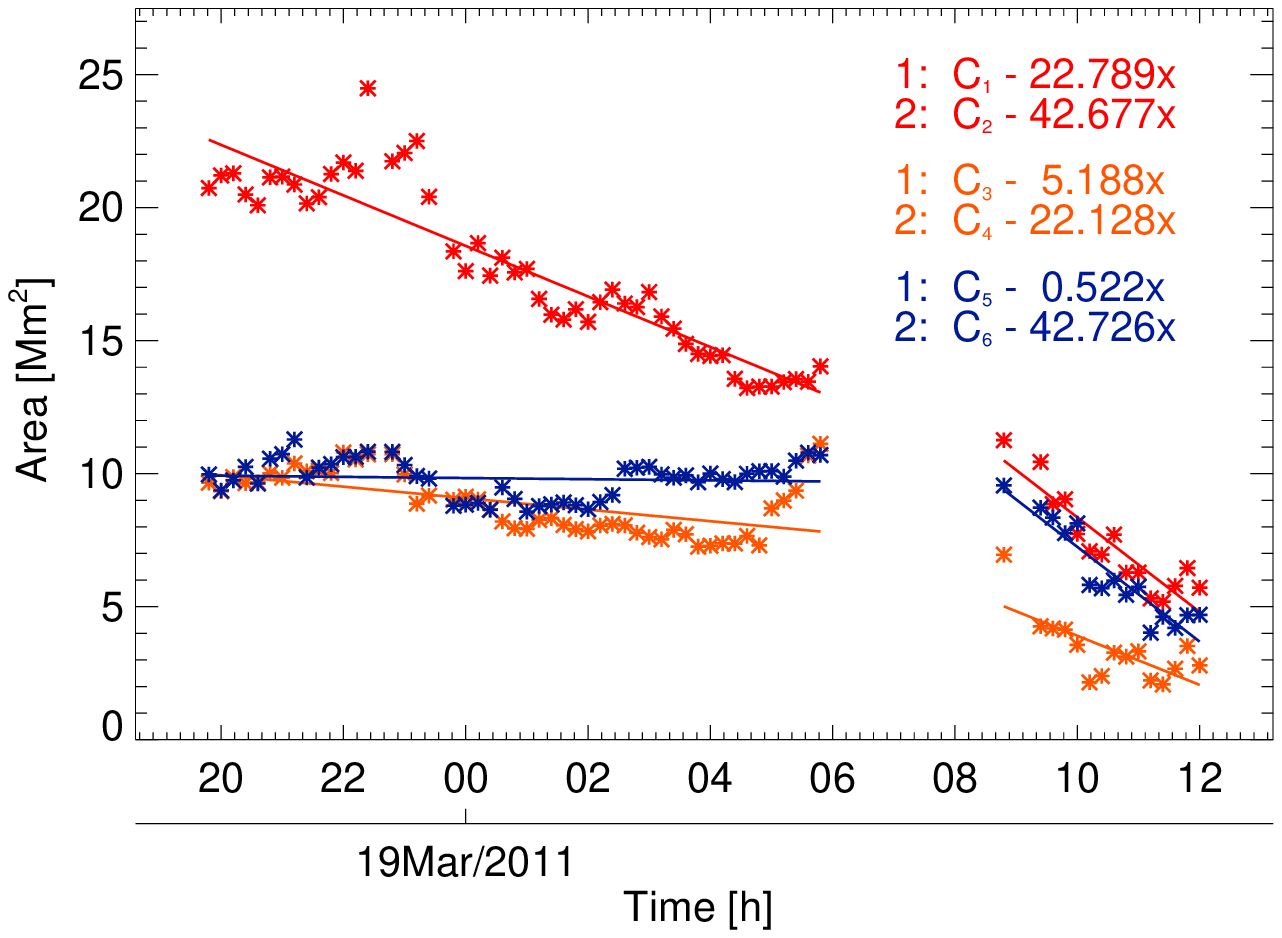}}
\caption{Temporal evolution of the areas of the pore encircled by both intensity and magnetic thresholds.\textit{ Top:} Comparison of the area of the pore encircled by the isocontours $I_c=0.55\,I_{QS}$ (red), $B=1921$~G (orange), and $B_\mathrm{ver}~=~1731$~G (blue). Vertical lines divide the evolutionary stages of the pore as in Fig.~\ref{fig:bver_evolution}.\textit{ Bottom:} Detail of the decaying stage. The straight lines are linear fits of the decay for each of the sub-periods (\textit{1, 2}) with the  continuous data available. 
}
\label{fig:areas_evolution}
\end{figure}

\section{Results} \label{sec:results}

We investigate the properties of the magnetic field on the boundary of an evolving pore. The analysis begins when the pore area, defined by $I_c\leq 0.55\,I_\mathrm{QS}$, exceeds five square pixels, which corresponds roughly to 0.72~Mm$^2$. The pore completely disintegrates into multiple areas smaller than 0.72~Mm$^2$ around 18:00~UT on March 19. From 12:00~UT on March 19, the pore is divided into multiple patches, which alternatively fulfil and do not fulfil the size requirement. Hence, in order to have a coherent analysis of the same magnetic structures during the whole evolution of the pore, we stopped the analysis at 12:00~UT. 

In each frame, we average the values of $\gamma$, $B$, and $B_\mathrm{ver}$ along the intensity isocontour and calculate the standard deviations of these magnetic parameters. We note that when the analysed pore splits into multiple parts, the presented values correspond to the average along all boundaries encircling areas larger than 0.72~Mm$^2$. Temporal evolutions of the averaged $B$, $\gamma$, and $B_\mathrm{ver}$ along the pore boundary are shown in Fig.~\ref{fig:bver_evolution}. It reveals that the temporal evolution of $B_\mathrm{ver}$ on the pore boundary behaves similarly to what has been observed on UP boundaries: $B_\mathrm{ver}$ on UP boundaries increases during the formation of the sunspot until it reaches a maximum stable value \citep{Jurcak_etal2015}, leading to a stage of stability characterised by a critical $B_\mathrm{ver}$ \citep{Jurcak2011} and followed by the decay of the sunspot, during which $B_\mathrm{ver}$ is weaker until the disappearance of the sunspot \citep{Benko_etal2018}.

Continuum intensity maps show a stage where the pore does not change significantly. It matches the period in which $B_\mathrm{ver}$ reaches its maximum value in Fig. \ref{fig:bver_evolution}. The combination of both the temporal evolution and continuum intensity maps suggests the existence of a stable stage in the lifetime of the pore with a constant $B_\mathrm{ver}$ that corresponds to the maximum value during its lifetime. Most importantly, this maximum $B_\mathrm{ver}$ value is comparable to the $B_\mathrm{ver}^\mathrm{crit}$ value found on UP boundaries of stable sunspots (see Sect.~\ref{discussion}).

\begin{figure*}[h]
{\includegraphics[width=\hsize]{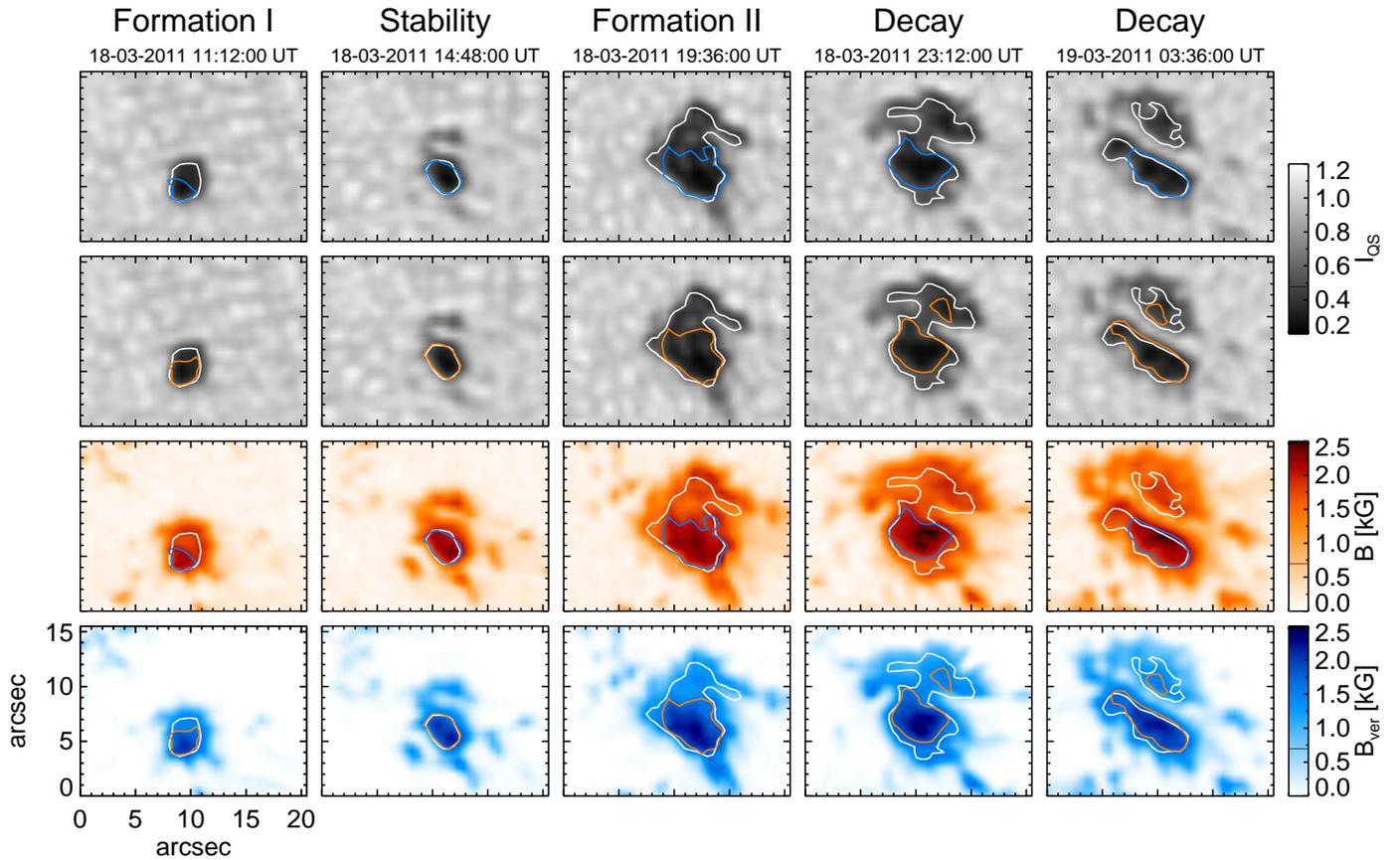}}
\caption{SDO/HMI samples of the evolutionary stages during the lifetime of the pore. \textit{Top row}: Intensity maps with intensity ($I_c=0.55\,I_\mathrm{QS}$; white) and magnetic field strength contours ($B=1921$~G; blue). \textit{Second row}: Intensity maps with intensity ($I_c=0.55\,I_\mathrm{QS}$; white) and $B_\mathrm{ver}$ contours ($B_\mathrm{ver}=1731$~G; orange). \textit{Third row}: Maps of magnetic field strength with contours identical to those in top row. \textit{Bottom row}: Maps of $B_\mathrm{ver}$ with contours identical to those in second row.}
\label{fig:Ic_maps}
\end{figure*}

 In Fig.~\ref{fig:areas_evolution}, we show the evolution of the pore area defined by the intensity threshold ($I_c\leq0.55\,I_\mathrm{QS}$; red symbols) next to the evolution of areas of the pore limited by magnetic thresholds ($B=1921$~G; orange symbols. $B_\mathrm{ver}=1731$~G; blue symbols). We define the evolutionary stages of the pore based on its size, morphology, and magnetic properties on its boundary. The phases, along with the definition of the magnetic thresholds, are described below.

\paragraph{Formation I.} After we started to analyse the pore on the solar surface at 09:24~UT, the pore area increased slowly from 1~Mm$^2$ to 2~Mm$^2$ until approximately 10:30~UT. This phase was accompanied by a strengthening of $B$ on its boundary along with an invariable $\gamma$, that is, the field became, on average, stronger on the pore boundary. We note that $B_\mathrm{ver}$ reached values around 1730~G during this short period of time. These values are comparable to the later identified critical value of $B_\mathrm{ver}$. This initial phase was followed by a sharp increase in the pore area (from 2~Mm$^2$ to 4~Mm$^2$) between 10:30~UT and 11:00~UT, when we found a local maximum of the pore size. During this rapid increase in the pore area, we found, on average, a weaker and more inclined field on the boundary, caused by the newly accumulated magnetic field. Between 11:30~UT and 13:30~UT, the pore area decreased slowly while, on average, $B$ increased and $\gamma$ decreased, resulting into an increase in $B_\mathrm{ver}$, on the boundary of the pore. We call this period of time the Formation~I phase, marked as For-I in Figs.~\ref{fig:bver_evolution} and~\ref{fig:areas_evolution}. In Fig.~\ref{fig:Ic_maps}, the left-most panels show the pore during this phase. The magnetic thresholds do not generally match the intensity boundary of the pore during this phase. 

\paragraph{Stability.} Between 13:30~UT and 16:30~UT, the pore size was not changing initially and then it slowly increased. We define this as the Stability stage. Within it, both $B$ and $B_\mathrm{ver}$ fluctuate around their maximum value over the pore lifetime. During this phase, the mean values weighted by their standard deviations are $B=1921$~G and $B_\mathrm{ver}=1731$~G. These values are used as thresholds of the magnetic parameters in our analysis based on the similar behaviour and strength of the $B_\mathrm{ver}$ found on the pore boundary and on UP boundaries, as explained above. As a result, the areas of the pore defined by the intensity and magnetic thresholds evolve consistently during the stable phase as shown in Fig.~\ref{fig:areas_evolution}. Accordingly, in Fig.~\ref{fig:Ic_maps}, the second column shows the pore in this phase, where we find a match between isocontours of intensity, $B$, and $B_\mathrm{ver}$.

\paragraph{Formation II.} The stable phase was terminated by a rapid growth of the pore caused by an accumulation of new flux in the northern part of the pore. This Formation~II phase lasted from 16:30~UT until 19:30~UT. During this time period, the pore size, defined by the intensity threshold, increased to 20~Mm$^2$. In the first 30~min, the pore expanded to an area with pre-existing magnetic field that strengthened and became more vertical, namely, it is also the areas outlined by $B_\mathrm{ver}=1731$~G and $B = 1921$~G that increased sharply (Fig.~\ref{fig:areas_evolution}, top plot). However, the newly gathered magnetic field is significantly weaker and more horizontal than the original magnetic field and leads to the decrease in the average $B$ and $B_\mathrm{ver}$ (and increase in $\gamma$) on the intensity boundary during the whole Formation~II period (Fig.~\ref{fig:bver_evolution}). During the whole period, the areas of the pore with $B_\mathrm{ver} > 1731$~G and $B > 1921$~G increased, but significantly more slowly than the area encircled by the isocontour $I_\mathrm{c}=0.55\,I_\mathrm{QS}$ (Fig.~\ref{fig:areas_evolution}, top plot). In Fig.~\ref{fig:Ic_maps}, we show the pore in the Formation~II phase in the third column. The magnetic thresholds show two distinctive regions: the original magnetic field with $B>1921$~G and $B_\mathrm{ver}>1731$~G in the southern part and the newly formed northern region with $B<1921$~G and $B_\mathrm{ver}<1731$~G. We note that the southern part of the pore boundary -- where there was no accumulation of new flux -- is mostly outlined by the magnetic thresholds.

\paragraph{Decay.} At $\sim$19:30~UT, the decay phase of the pore began. Initially, between 19:30~UT and 23:00~UT, the pore size fluctuated around 21~Mm$^2$ and the areas with $B_\mathrm{ver} > 1731$~G and $B>1921$~G slightly increased. Afterwards, between 23:00~UT and 06:00~UT, the pore decreased in size while regions with $B>1921$~G tend to diminish. However, during this period, areas of the pore with $B_\mathrm{ver} > 1731$~G fluctuated around 10~Mm$^2$. Then both the areas defined by intensity and magnetic thresholds decreased. Decay rates were studied separately during the two sub-periods with the  continuous data available (bottom plot in Fig.~\ref{fig:areas_evolution}). We note that during the whole decay process, we observed slight fluctuations of the pore areas as defined both by intensity and magnetic thresholds. These fluctuations are partly caused by relatively fast evolution of outermost structures of the pore with intensities comparable to the intensity threshold and partly caused by the size increase due to occasional small accumulations of magnetic flux.

The first sub-period spans from 18~March, 19:30~UT to 19~March, 06:00~UT. During this interval, the pore defined by the intensity isocontour decays at a rate of 0.95~$\mathrm{Mm^2/h}$. Regions with $B>1921$~G decay at a rate of 0.22~$\mathrm{Mm^2/h}$ on average. This implies that regions with $B<1921$~G decay at a rate of 0.73~$\mathrm{Mm^2/h}$, which is 3.4~times faster than regions with a stronger total magnetic field. Regions with $B_\mathrm{ver}>1731$~G decay at a rate of 0.02~$\mathrm{Mm^2/h}$. This implies, as observed in Fig.~\ref{fig:areas_evolution}, that only regions with a weak vertical magnetic field disappear from the solar surface while regions with a strong vertical magnetic field remain almost invariant.
 Short-term fluctuations of the pore size as defined by intensity and magnetic thresholds are small compared to the overall trend observed in this sub-period. At the end of the first sub-period, we observe a newly accumulated magnetic flux in the northern region that has more significant impact on the size of regions defined by the total magnetic field strength threshold. This indicates an accumulation of a strong but inclined magnetic field.

The second sub-period spans from 19~March, 08:48~UT to 12:00~UT. During this interval, the pore size defined by the intensity isocontour decays at a rate of 1.78~$\mathrm{Mm^2/h}$ and it is an identical decay rate as for regions with $B_\mathrm{ver}>1731$~G, while regions with $B>1921$~G decay at a rate of 0.922~$\mathrm{Mm^2/h}$. During this sub-period, the patches of the pore defined by the intensity and $B_\mathrm{ver}$ thresholds are nearly identical, and therefore we observe $B_\mathrm{ver}$ values close to 1731~G during this period (see Fig.~\ref{fig:bver_evolution}). It means that the dissipation of the pore is strictly connected to the disappearance of strong vertical fields. The flux tube is mostly vertical and the total magnetic field strength is, in general, weaker than in previous stages. Fluctuations of the pore regions defined by intensity and magnetic thresholds are significant compared to the mean values during the second sub-period of the decay. Towards the end of the lifetime of the pore, the pore is split into multiple segments. These segments alternately fulfil or do not fulfil the conditions to be included in the analysis and, thus, they cause rapid changes in the pore areas defined by the intensity and magnetic thresholds, which significantly influences the evolving properties and the decay rates, therefore, we have not included this last sub-period in the analysis.

In Fig.~\ref{fig:Ic_maps}, the fourth and fifth columns show the pore in the Decay phase. We observe a division of the pore by a light bridge through a region with weak magnetic field. A comparison of the two frames also indicates that the area of the pore with $B_\mathrm{ver}~<~1731$~G ($B~<~1921$~G) is more unstable. We also want to note that during the Decay phase, we find a better match between $I_c$ and $B_\mathrm{ver}$ contours, rather than between $I_c$ and $B$ contours on the southern segment of the pore, which is the most stable region. 

\section{Discussion and conclusions}\label{sec_conclusions}
\label{discussion}

In this work, we analyse the magnetic properties on the boundary of an evolving pore with the aim of investigating the role of the vertical component of the magnetic field on the pore stability. More specifically, we study the similarity between the umbra-penumbra (UP) boundaries in sunspots and pore-quiet Sun boundaries in terms of $B_\mathrm{ver}^\mathrm{crit}$ values. The pore stability refers to the effective damping of more vigorous magneto-convective motions and, conversely, the pore instability refers to the ineffective attenuation of vigorous magneto-convective motions within the pore, leading to its dissipation. We use 113 observations of a pore in active region NOAA~11175 taken by SDO/HMI that cover $\sim\,$26.5 hours of the pore evolution. A morphological characterisation of the evolution of the pore, determined from the development of the pore area (Fig.~\ref{fig:areas_evolution}) and from the behaviour of the averaged magnetic parameters along the pore boundary (Fig.~\ref{fig:bver_evolution}), lead us to define four evolving phases: a first formation period (Formation I), a stable period (Stability), a second formation period (Formation II), and a disintegration period (Decay).

During the first formation phase, we find that both $B$ and $B_\mathrm{ver}$ are generally increasing on the visual boundary of the pore defined at $I_c=0.55\,I_\mathrm{QS}$. In the later phase of the Formation I period, the pore area decreases until the maximum values of $B$ and $B_\mathrm{ver}$ are reached on the boundary (Fig.~\ref{fig:bver_evolution}). During the Stability phase that follows, the pore boundaries defined by $0.55\,I_\mathrm{QS}$, $B = 1921$~G and $B_\mathrm{ver} = 1731$~G are nearly identical and encircle the same area (Fig.~\ref{fig:areas_evolution}). This stability is disrupted by the accumulation of new magnetic flux of weaker and more horizontal field that causes a sharp increase in the pore size and the subsequent decrease in $B$ and $B_\mathrm{ver}$ on the pore boundary. The pore does not stabilise again and starts to decay immediately after the supply of new magnetic flux is depleted. 

The mean value of $B_\mathrm{ver}$ found on the pore boundary during the Stability phase is 1731~G. This value is comparable to 1639~G, $B_\mathrm{ver}$ found by \citet{Schmassmann_etal2018} on the UP boundary of a long-lived sunspot, where the authors analysed standard SDO/HMI data at 53\% $I_\mathrm{QS}$. The dissimilarity can be attributed to the use of deconvolved data, which should lead to stronger $B_\mathrm{ver}$ values. The similarity of the $B_\mathrm{ver}$ value on the pore boundary with those found on UP boundaries of stable sunspots \citep[1867~G, 1693~G, 1787~G by][respectively]{Jurcak_etal2018, Schmassmann_etal2018, Lindner:2020} indicates that it is also in pores that the magneto-convection is effectively and stably hindered by the vertical component of the magnetic field. In the analysed pore, we find this critical value to be $B_\mathrm{ver}^\mathrm{crit}$ $\sim$1730~G.

Furthermore, in the studied pore, we find that the same applies for a magnetic field strength of 1921~G, the mean value on the pore boundary during the Stability phase. During this phase, the isocontours of intensity and $B$ are in equally good agreement as isocontours of intensity and $B_\mathrm{ver}$. Also, the areas encircled by isocontours of $B=1921$~G and $B_\mathrm{ver}=1731$~G, shown in Fig.~\ref{fig:areas_evolution}, exhibit a comparable evolution. 

During the Decay phase, we initially observe a much faster disappearance of the pore areas with $B_\mathrm{ver}<1731$~G, namely, granular magneto-convection takes over pore areas with $B_\mathrm{ver} < B_\mathrm{ver}^\mathrm{crit}$. This provides us with a further indication of the important role of $B_\mathrm{ver}^\mathrm{crit}$ in hindering magneto-convection, as stated by the Jur\v c\'ak criterion for stable sunspots. The decay process takes hours and is thus comparable to other case studies. For example, \cite{Jurcak_etal2017} analysed the 12-hour formation of a penumbra at the expense of a pore, which had $B_\mathrm{ver} < B_\mathrm{ver}^\mathrm{crit}$ over the whole process; and \citet{Benko_etal2018} described the decay of a sunspot umbra that spans over several days. We do not know why the decay process is not comparable to the granular life-time or life-time of penumbral filaments.

For the first time, we try to directly estimate, the stabilising role of the critical value of the vertical component of the magnetic field by comparing the decay rates of the areas that have $B_\mathrm{ver}>1731$~G and $B_\mathrm{ver}<1731$~G. During the initial sub-period of the pore's decay, we find that areas with $B_\mathrm{ver}>1731$~G do not decay, while areas with $B_\mathrm{ver}<1731$~G progressively disappear until, after the gap in the observations, the areas of the pore defined by the intensity and $B_\mathrm{ver}$ thresholds are nearly identical. Therefore, we observe an increase in the $B_\mathrm{ver}$ value averaged over the intensity boundary of the pore towards the end of the analysed period (see Fig.~\ref{fig:bver_evolution}).

 As pointed out earlier in this work, we can also use $B$ as a determining parameter for stability in the studied pore. Areas with $B>1921$~G decay 3.4 times slower than pore areas with $B<1921$~G. We note that during the second sub-period of the pore's decay, the decay rates for areas with $B_\mathrm{ver}>1731$~G and $B>1921$~G increased significantly, and regions with $B_\mathrm{ver}>1731$~G decay at the same rate as areas defined by the intensity threshold. This means that the pore decays at a fast rate while having an intense vertical magnetic field, and we do not know the reason for this behaviour.

This case study highlights the importance of the $B_\mathrm{ver}^\mathrm{crit}$ value found for sunspots to the stability of pores as well. However, an analysis of a larger sample of pores is necessary in order to investigate if the found $B_\mathrm{ver}^\mathrm{crit} \sim 1730$~G is a unifying parameter for stable boundaries of all pores and if the same unifying parameter could also be $B^\mathrm{crit}~\sim~1920$~G. Statistical analyses of pore lifetimes and decay rates with respect to their magnetic properties is also necessary in order to investigate the stabilising effect of $B_\mathrm{ver}^\mathrm{crit}$.

\begin{acknowledgements}
      We would like to thank the referee for the valuable comments that helped to improve this paper. We also would like to thank A. Norton for providing us deconvolved full-disc vector field maps. This work was supported by project 204119 from the Grant Agency of Charles University and by the Czech Science Foundation grant project 18-06319S. The data is courtesy of NASA/SDO and the HMI science teams. Hinode is a Japanese mission developed and launched by ISAS/JAXA, with NAOJ as domestic partner and NASA and STFC (UK) as international partners. It is operated by these agencies in co-operation with ESA and NSC (Norway).
\end{acknowledgements}

\bibliographystyle{aa}
\bibliography{biblio}

\end{document}